\documentclass{cernrep} 
\usepackage{texnames}
\usepackage[T1]{fontenc}
\usepackage[bookmarks, colorlinks=true, linktoc=page, pdftex, linkcolor=black, citecolor=black, urlcolor=blue]{hyperref}
\def\mus{\mu{\rm{s}}}
\def\sec{{\rm{s}}}
\def\ab{\rm{ab}}
\def\fb{\rm{fb}}
\def\m{\rm{m}}
\def\cm{\rm{cm}}
\def\mm{\rm{mm}}
\def\km{\rm{km}}
\def\mim{\mu{\rm{m}}}
\def\Hz{\rm{Hz}}
\def\rad{\rm{rad}}
\def\MW{\rm{MW}}
\def\MV{\rm{MV}}
\begin{document}
\vspace*{10pt}
\begin{flushright}
{\bf{Input to the European Particle Physics Strategy Update}} \\
\end{flushright}

\begin{center}
\vspace*{15mm}

\vspace{0.8cm}
{\bf \Large Muon Colliders}\\

\vspace{0.9cm}

{\bf The Muon Collider Working Group}\\
{Jean~Pierre~Delahaye$^1$, Marcella~Diemoz$^2$, Ken~Long$^3$, Bruno~Mansouli\'{e}$^4$, Nadia~Pastrone$^5$~(chair), Lenny~Rivkin$^6$, Daniel~Schulte$^1$, Alexander~Skrinsky$^7$, Andrea~Wulzer$^{1,8}$}

\vspace{.5cm}
$^1$ CERN, Geneva, Switzerland\\
$^2$ INFN Sezione di Roma, Roma, Italy\\
$^3$ Imperial College, London, United Kingdom\\
$^4$ CEA, IRFU, France\\
$^5$ INFN Sezione di Torino, Torino, Italy\\
$^6$ EPFL and PSI, Switzerland\\
$^7$ BINP, Russia\\
$^8$ LPTP, EPFL, Switzerland and University of Padova, Italy\\
\end{center}
\vspace{.5cm}
\medskip
\noindent Muon colliders have a great potential for high-energy physics. 
They can offer collisions of point-like particles at very high energies, since muons can be accelerated in a ring without limitation from synchrotron radiation. However, the need for high luminosity faces technical challenges which arise from the short muon lifetime at rest and the difficulty of producing large numbers of muons in bunches with small emittance.  Addressing these challenges requires the development of innovative concepts and demanding technologies.

\noindent  The document summarizes the work done, the progress achieved and new recent ideas on muon colliders. A set of further studies and actions is also identified to advance in the field. Finally, a set of recommendations is listed in order to make the muon technology mature enough to be favourably considered as a candidate for high-energy facilities in the future.

\vfill
\vspace{1pt}
\noindent{\bf{Contact:}} Nadia Pastrone, nadia.pastrone@cern.ch\\
\noindent{\bf{Webpage:}}  https://muoncollider.web.cern.ch\\
\vspace{10pt}

 \newpage
\section{Introduction}
 
The quest for discovery in particle physics has always required experiments at the highest possible energies. Novel techniques have been developed to accelerate beams to very high energy, to bring these beams into collision and to detect the particles that these collisions create. The successful discovery of the Higgs boson and the search for new particles and forces at the LHC rests on this tradition. Further advances in technique will be required to create a facility capable of continuing the search for phenomena beyond the Standard Model into an extended energy domain. Such a facility will be required to deliver collisions at centre-of-mass energies in excess of few TeV and be capable of measuring the collision products with precision, in order to explore new physics both directly and indirectly. These features would be needed also to characterise possible LHC or HL-LHC discoveries. The fundamental nature of leptons makes lepton-antilepton colliders the ideal candidate to serve at the energy frontier. Muon colliders offer great potential for discovery in the multi-TeV energy range. The large muon mass, $207$ times that of the electron, suppresses synchrotron radiation by a factor of $10^9$ compared with electron beams of the same energy. Rings can therefore be used to accelerate muon beams efficiently and to bring them into collision repeatedly. Also, more than one experiment can be served simultaneously.

The physics reach of the muon collider is extended over that of a proton-proton collider of the same energy since all of the beam energy is available for the hard collision, compared to the fraction of the proton-beam energy carried by the colliding partons. A $14$~TeV muon collider provides an effective energy reach similar to that of the 100 TeV FCC, and potentially the c.o.m. energy in colliding muons can go well beyond. Furthermore a dedicated muon collider is the only lepton-antilepton collider option able to scan the Higgs resonance and to precisely measure its mass and width. A muon collider is therefore ideal to search for and/or study new physics and for resolving narrow resonances both as a precision facility and/or as an exploratory machine as is developed in section~\ref{sec:physics}. 

A muon collider has the potential to provide excellent performance at reasonable power consumption. A figure of merit for energy-frontier facilities may be defined as the luminosity per unit power. This figure of merit appears to be substantially larger for the muon collider than for any other lepton collider technology in the multi-TeV range \cite{Boscolo:2018ytm}.

The main challenges of a muon collider design are those arising from the short muon lifetime, which is $2.2~\mus$ at rest, and the difficulty of producing large numbers of muons in bunches with small emittance. This requires the development of demanding technologies and innovative concepts, as presented in section~\ref{sec:machine}. The beam background from the decay of the muons impacts both machine and detector design (see section~\ref{sec:background}).  

In addition, muon decays offer a window on another important area of physics by generating electron and muon neutrinos with a well-known flux and energy spectrum. These neutrinos can serve as the source for a Neutrino Factory (NF \cite{Choubey:2011zzq}), which would constitute the ideal complement  \cite{Huber:2014nga} to Long Base Line Facilities. 

\section{Physics}\label{sec:physics}

\begin{figure}[t]
\includegraphics[width=.45\linewidth]{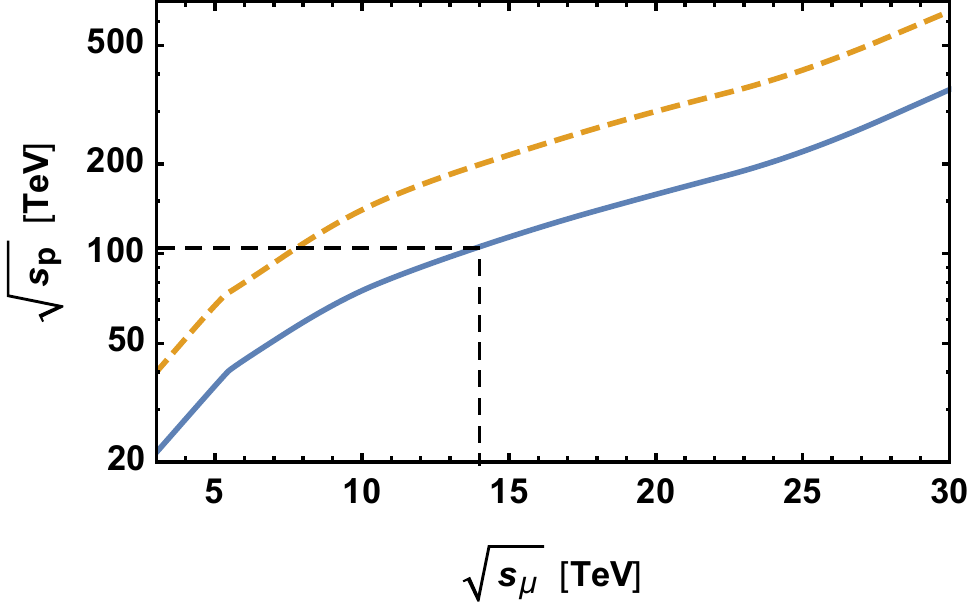}\hspace{30pt}
\raisebox{.006\linewidth}{\includegraphics[width=.42\linewidth]{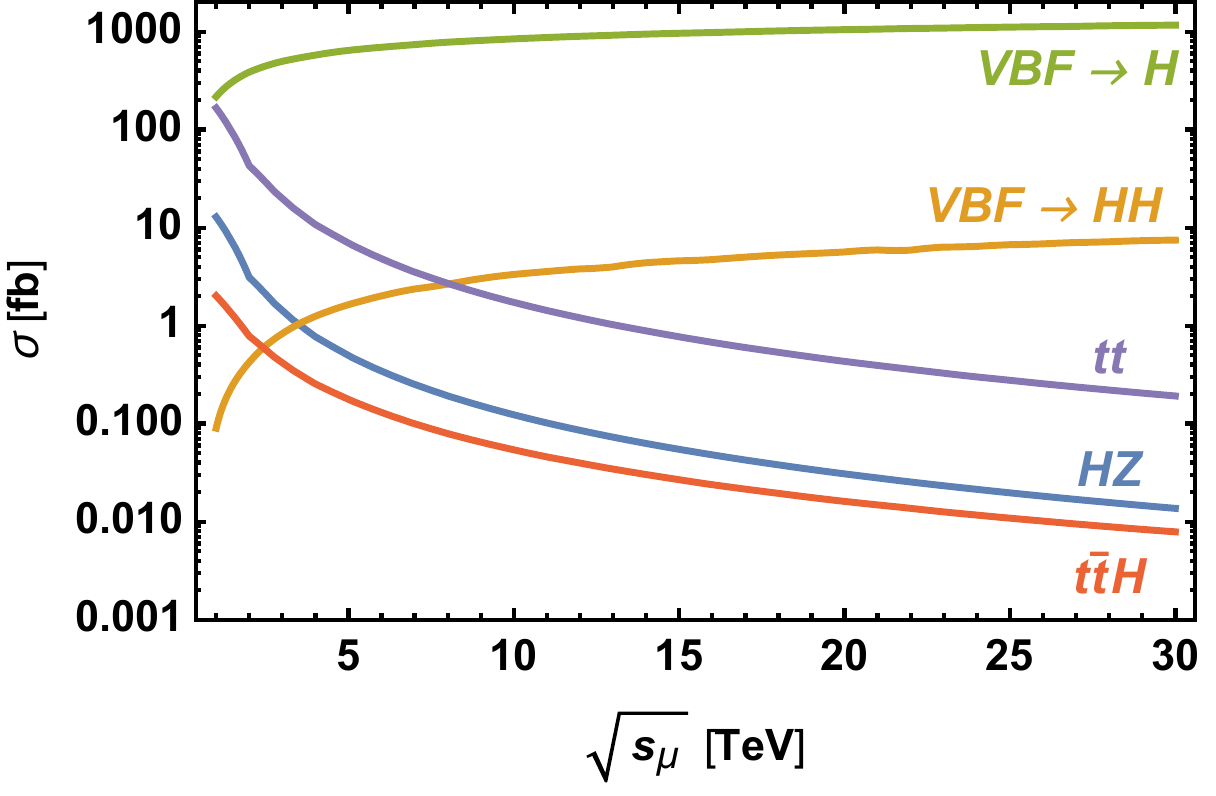}}
\caption{Left panel: the energy at which the proton collider cross-section equals that of a muon collider. The dashed line assumes comparable Feynman amplitudes for the muon and the proton production processes. A factor of ten enhancement of the proton production amplitude squared, possibly due to QCD production, is considered in the continuous line. Right panel: Higgs and top-quark production cross-sections at high energy lepton colliders.}
\label{fig:muvsp}
\end{figure}

Circular muon colliders have the potential to reach center-of-mass energies of tens of TeV. Since such high collision energies are currently achievable only with proton-proton machines, it is natural to start an assessment of the muon collider physics potential from a comparison with the hadronic option. The obvious advantage in colliding muons rather than protons is that the muon collider center of mass energy, $\sqrt{s}_\mu$, is entirely available to produce short-distance reactions. At a proton collider, instead, the relevant interactions occur between the proton constituents, which carry a small fraction of the collider energy, $\sqrt{s}_p$. This can be illustrated by considering the pair production of heavy particles with mass $M$ approximately equal to half the muon collider energy.\footnote{The estimate that follows generically applies to any $2\rightarrow2$ reaction with a high threshold in energy.} The cross-sections scales as $1/s$ at both colliders, but at a proton collider the cross-section is to be multiplied by an $M/\sqrt{s}_p$-dependent suppression factor due to the steeply-falling parton luminosities. Equal muon and proton collider cross-sections are thus obtained for $\sqrt{s}_\mu\ll \sqrt{s}_p$, as shown on the left panel of Figure~\ref{fig:muvsp}.

Naively, one would expect the lower background level could be another advantage of the muon collider relative to hadronic machines. However it is unclear to what extent this is the case because of the large beam background from the decay of the muons, discussed in section~\ref{sec:background}.  
 
Figure~\ref{fig:muvsp} suggests that a $14$~TeV muon collider with sufficient luminosity might be very effective as a {\bf{direct exploration}} machine, with a physics motivation and potential similar to that of a $100$~TeV proton-proton collider \cite{Arkani-Hamed:2015vfh}.  Although detailed analyses are not yet available, it is expected that a future energy frontier muon collider could make decisive progress on several beyond-the-SM questions, and to be conclusive on some of these questions. By exploiting the very large vector-boson fusion (VBF) cross-section, a muon collider could search extensively for new particles coupled with the Higgs boson, possibly related to electroweak baryogenesis \cite{Buttazzo:2018qqp}. It might also discover Higgsinos or other heavy WIMP dark matter scenarios \cite{DiLuzio:2018jwd}. In this context, it is important to remark that motivated ``minimal'' WIMP dark matter candidates might have a mass of up to $16$~TeV. Generic electroweak-charged particle with easily identifiable decay products up to a mass of several TeV can be searched for. Relevant benchmarks are the (coloured) top partners related with naturalness, which should be present at this high mass even in elusive ``neutral naturalness'' scenarios. 

The ability to perform measurements, which {\bf{probe New Physics indirectly}} \footnote{Precision would also allow the characterization of newly discovered particles.}, is another important goal of future collider projects. The high energy of a muon collider could also be beneficial from this viewpoint, in two ways. First, indirect New Physics effects are enhanced at high energy, so that they can show up even in relatively inaccurate measurements. This is the mechanism by which the $3$~TeV CLIC might be able to probe the Higgs compositeness scale above $10$~TeV (or a weakly-coupled $Z^\prime$ up to $30$~TeV) with  di-fermion and di-boson measurements at the $1\%$ level \cite{deBlas:2018mhx}, while an exquisite precision of $10^{-4}/10^{-5}$ would be needed to achieve the same goal with low-energy (e.g., $Z$-pole) observables. At a $30$~TeV muon collider, with suitably scaled luminosity, the reach would increase by a factor of $10$. The second important aspect is that some of the key processes for Higgs physics, namely those initiated by the vector boson fusion (see the right panel of Figure~\ref{fig:muvsp}), have very large cross-sections. For instance with an integrated luminosity of $10~\ab^{-1}$, a $10$~TeV muon collider would produce $8$ million Higgs bosons, with \mbox{$30$'$000$} of them by the pair production mechanism that is sensitive to the trilinear Higgs coupling. While further study is required, especially in view of the significant level of machine background that is expected at a muon collider, these numbers might allow a satisfactory program of Higgs couplings determination.

A detailed assessment of the muon collider luminosity requirements will result from a comprehensive investigation of the physics potential, which is not yet available. However a simple and robust estimate of the minimal useful luminosity can be readily obtained as follows. Reactions induced by electroweak interactions, with cross-section of order
\begin{equation}\label{eq:XS}
\sigma=\left(
\frac{10\,{\rm{TeV}}}{\sqrt{s}_\mu}\right)^2\cdot1\,\fb\,,
\end{equation}
are the benchmark processes at a muon collider. This is the cross-section estimate for both the production of new particles with a mass of the order of (or much below) the collider energy, or large-$P_T$  $2\rightarrow2$ SM processes.  The number of events corresponding to Eq.~(\ref{eq:XS}), normalized to $5$~years run and assuming \mbox{$10^7\,\sec/$year} operation, is
\begin{equation}
N=
\frac{\rm{time}}{5\,{\rm{yrs}}}
\left(
\frac{10\,{\rm{TeV}}}{\sqrt{s}_\mu}\right)^2\frac{\rm{luminosity}}{10^{34}{\rm{cm}}^{-2}\rm{s}^{-1}}
\cdot
500
\,.
\end{equation}
Collecting $100$ events might be sufficient to discover new particles with easily identifiable decay products, such as Stops and Top Partners related with Naturalness. An instantaneous luminosity of $2\cdot10^{33}{\rm{cm}}^{-2}\rm{s}^{-1}$, at $10$~TeV, would be sufficient to probe these particles up to the collider reach. Ten thousands events would instead be needed to aim at percent-level measurements of electroweak SM processes at high invariant mass, allowing to probe hundreds of TeV New Physics scales indirectly as previously mentioned. In this case the luminosity requirement becomes:
\begin{equation}
L\gtrsim\frac{5\,{\rm{years}}}{\rm{time}}
\left(
\frac{\sqrt{s}_\mu}{10\,{\rm{TeV}}}
\right)^2
2\cdot10^{35}{\rm{cm}}^{-2}{\rm{s}}^{-1}\,.
\end{equation}
If extrapolated down to $3$~TeV, this formula corresponds to a luminosity of around $2\cdot10^{34}{\rm{cm}}^{-2}{\rm{s}}^{-1}$, which is a factor of 3 less than that of the 3 TeV CLIC~\cite{Aicheler:2012bya}. However thanks to the absence of beamstrahlung this luminosity is entirely available for high-energy reactions, and furthermore two interaction points are foreseen at the muon collider.  At $10$~TeV, the formula corresponds to $10\,\ab^{-1}$ integrated luminosity in $5$ years, which is the one assumed above in the estimate of the total number of produced Higgs bosons.

\subsubsection*{Physics potential along the way}
A vigorous and ambitious R\&D program is needed to assess the feasibility of a tens-of-TeV's muon collider. Therefore it is important to investigate the physics potential of smaller-scale machines that might be built along the way as technology demonstrators. Starting from medium energy, the first option to be considered is a muon collider operating around the top production threshold ($\sim400$~GeV). This could have the same potential as the  CLIC Stage~1 \cite{CLIC:2016zwp} in terms of top \cite{Abramowicz:2018rjq} and Higgs \cite{Abramowicz:2016zbo} physics, provided a comparable luminosity (of the order of $10^{34}{\rm{cm}}^{-2}{\rm{s}}^{-1}$) is obtained. The physics case becomes less clear at lower energies, where the muon collider luminosity is expected to be not comparable with that of circular $e^+e^-$ machines. A remarkable exception is of course a muon collider operating at the Higgs pole, which might study the Higgs boson line-shape. See Ref.s~\cite{Janot} and \cite{Blondel} for a recent study, which assumes $4~\fb^{-1}$ integrated luminosity and $3\cdot 10^{-5}$ relative energy spread of the muon beams. Comparisons with updated projections from HL-LHC shows that the impact on the Higgs coupling will be limited, but significant progress is possible in the measurement of the Higgs mass and in the model-independent determination of the Higgs width. In particular it is worth emphasizing that the Higgs mass is a fundamental input parameter of the SM, which would be measured at one part per million. On the other hand, the practical impact of such an astonishingly precise measurement is not fully clear yet.

\section{Machine Design}\label{sec:machine}

\subsection{Introduction}

\begin{figure}[t]
\centering
\includegraphics[width=.65\linewidth]{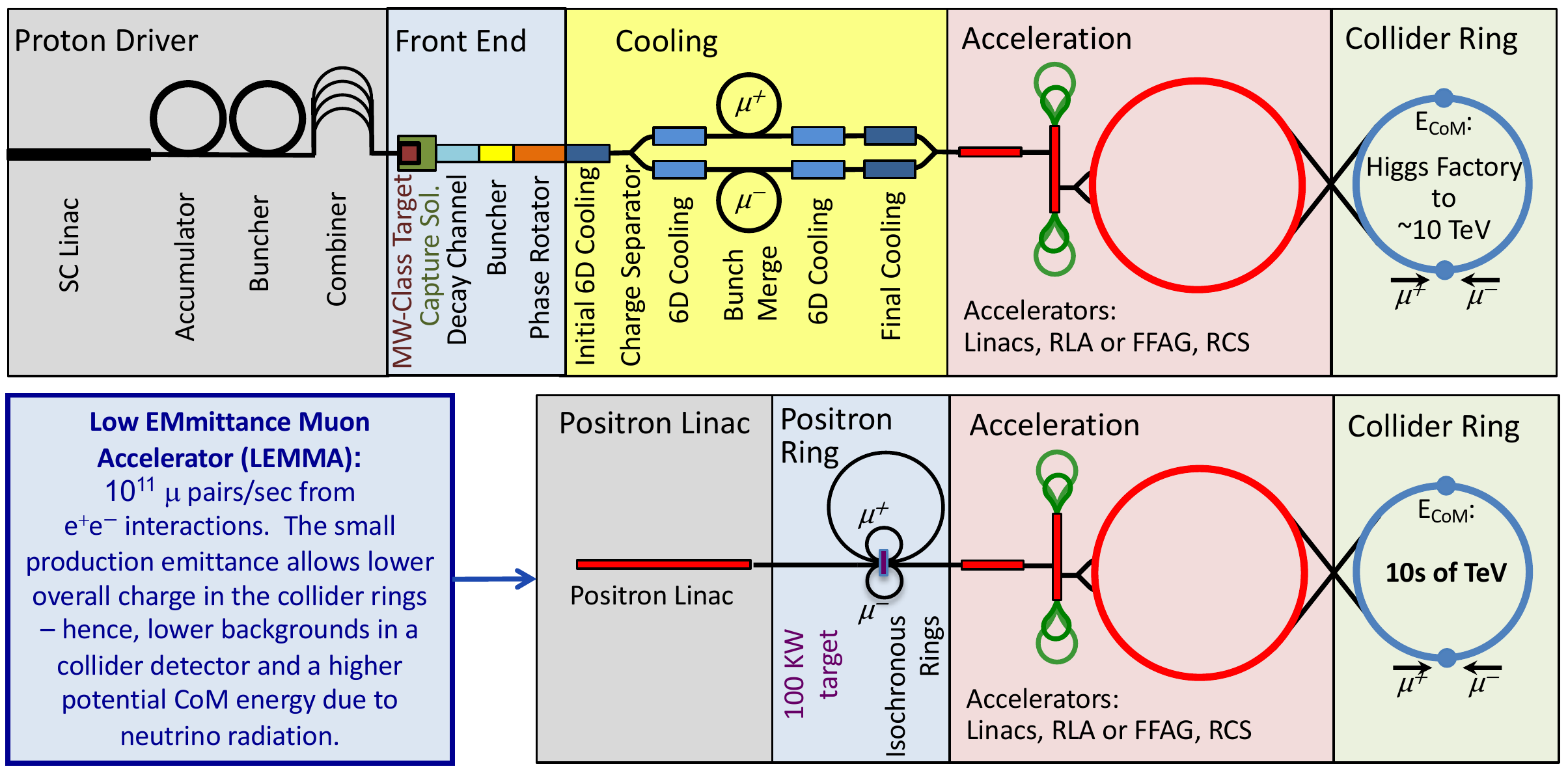}
\caption{Schematic layouts of Muon Collider complexes based on the proton driver scheme and on the low emittance positron driver scheme emphasizing synergies.}
\label{fig:scheme}
\end{figure}

Because of their great potential and critical challenges, muon-based facility concepts have been developed for more than three decades. The present status of the various options considered and the necessary R\&D to address their feasibility is summarized in Ref.~\cite{Boscolo:2018ytm}. Their basic layouts are shown in Figure~\ref{fig:scheme}, emphasizing synergies. The idea of muon colliders was first introduced in the early 1980's \cite{Skrinsky:1981ht, Neuffer:1983xya} and further developed by a series of world-wide collaborations \cite{Palmer,Palmer:2014nza} culminating in creation of the US Muon Accelerator Program (MAP) \cite{MAP} in 2011. MAP developed the concepts of a proton driver scheme and addressed the feasibility of the novel technologies required for Muon Colliders and Neutrino Factories \cite{JINST}.  In the scheme (see section~\ref{sec:prds}), the muons are generated as tertiary particles in the decays of the pions created by an intense proton beam interacting a heavy material target. In order to achieve high luminosity in the collider, the resulting initial low energy muon beam with short lifetime, with large transverse and longitudinal emittances, has to be cooled by five orders of magnitude in the six-dimensional phase-space and rapidly accelerated to minimize the decrease of the intensity due to muon decays.   

A novel approach of the Low Emittance Muon Accelerator (LEMMA) based on muon pair production with a positron beam impinging on electrons at rest in a target \cite{Antonelli:2013mmk} was recently proposed and is now under conceptual study \cite{Antonelli:2015nla}. The corresponding positron driver scheme is described in section~\ref{sec:pods}. The muons produced in the $e^+e^-$ interactions close to threshold are constrained into a small phase-space region, effectively producing a muon beam with very small transverse emittances \cite{Boscolo:2017xmn}, comparable to those typically obtained in electron beams without necessitating any cooling. These muon pairs are produced with an average energy of $22$~GeV corresponding to an average laboratory lifetime of $\sim500\,\mus$, which mitigates the intensity losses by muon decay and eases the acceleration scheme. Potentially high luminosity could be reached with relatively small muon fluxes, reducing background and activation problems due to high energy muon decays, and thus mitigating the on-site neutrino radiation issue. Consequently, the LEMMA scheme, although not appropriate for a Higgs Factory due to a too large beam energy spread, is very attractive for a collider in the multi-TeV range, extending the energy reach of muon colliders which can be limited by neutrino radiation. 

\subsection{Proton driver scheme}\label{sec:prds}

\subsubsection{Design status}
In the proton driver scheme \cite{Palmer:2014nza, MAP} muons are produced as tertiary particles from decay of pions created by a high-power proton beam impinging a high Z material target. The majority of the produced pions have momenta of a few hundred \mbox{MeV$/c$}, with a large momentum spread and large transverse momentum components. Hence, the daughter muons are produced at low energy within a large longitudinal and transverse phase-space. This initial muon population must be confined transversely, captured longitudinally, and have its phase-space manipulated to fit within the acceptance of an accelerator.  These beam manipulations must be done quickly, before the muons decay.

A schematic layout of a proton driven muon collider facility is sketched in Figure~\ref{fig:scheme}. The main parameters of the enabled facilities are summarized in Table~\ref{tab:pars}.

\begin{table}[t]\caption{Main parameters of the proton driver muon facilities}\label{tab:pars}
\begin{tabular}{lccccc}
\hline
{\bf{Parameter}} & {\bf{Units}} & {\bf{Higgs}} & \ & {\bf{Multi-TeV}} & \  \\
CoM Energy & TeV & $0.126$ & $1.5$ & $3.0$ & $6.0$\\
Avg. Luminosity & $10^{34}\cm^{-2}\sec^{-1}$ & $0.008$ & $1.25$ & $4.4$ & $12$\\
Beam Energy Spread & $\%$ & $0.004$ & $0.1$ & $0.1$ & $0.1$\\
Higgs Production$/10^7$ sec & \ & $13$'$500$ & $37$'$500$ & $200$'$000$ & $820$'$000$\\
Circumference & $\km$ & $0.3$ & $2.5$ & $4.5$ & $6$ \\
No. of IP's & \ & $1$ & $2$ & $2$ &$2$ \\ 
Repetition Rate & $\Hz$ & $15$ & $15$ & $12$ & $6$ \\
$\beta^*_{x,y}$ & $\cm$ & $1.7$ & $1$ & $0.5$ & $0.25$ \\
No. muons/bunch & $10^{12}$ & $4$ & $2$ & $2$ & $2$ \\
Norm. Trans. Emittance, $\varepsilon_{\rm{TN}}$ & $\mim$-\rad & $200$ & $25$ & $25$ & $25$ \\
Norm. Long. Emittance, $\varepsilon_{\rm{LN}}$ & $\mim$-\rad & $1.5$ & $70$ & $70$ & $70$ \\
Bunch Length, $\sigma_{\rm{S}}$ & $\cm$ & 6.3 & 1 & 0.5 & 0.2 \\
Proton Driver Power & $\MW$ & $4$ & $4$ & $4$ & $1.6$ \\
Wall Plug Power & $\MW$ & $200$ & $216$ & $230$ & $270$ \\
\hline
\end{tabular}
\end{table}

The functional elements of the muon beam generation and acceleration systems are:
\begin{itemize}
\item a proton driver producing a high-power multi-GeV, multi-$\MW$ bunched $H^- beam$,
\item a buncher made of an accumulator and a compressor that forms intense and short proton bunches, 
\item a pion production target in a heavily shielded enclosure able to withstand the high proton beam power, which is inserted in a high field solenoid to capture the pions and guide them into a decay channel,
\item a front-end made of a solenoid decay channel equipped with RF cavities that captures the muons longitudinally into a bunch train, and then applies a time-dependent acceleration that increases the energy of the slower (low-energy) bunches and decreases the energy of the faster (high-energy) bunches,
\item  an ``initial'' cooling channel that uses a moderate amount of ionization cooling to reduce the 6D phase space occupied by the beam by a factor of $50$ ($5$ in each transverse plane and $2$ in the longitudinal plane), so that it fits within the acceptance of the first acceleration stage. For high luminosity collider applications, further ionization cooling stages are necessary to reduce the 6D phase space occupied by the beam by up to five orders of magnitude,
\item the beam is then accelerated by a series of fast acceleration stages such as Recirculating Linacs Accelerators (RLA) or Fixed Field Alternating Gradient (FFAG) and Rapid Cycling Synchrotron (RCS) to take the muon beams to the relevant energy before injection in the muon collider Ring.
\end{itemize}

\subsubsection{R\&D}

The MAP R\&D program (2011-2018) addressed many issues toward technical and design feasibility of a muon based neutrino factory or collider \cite{JINST} . Significant R\&D progress, also summarized in \cite{Boscolo:2018ytm}, was achieved.

\noindent\underline{\it{Operation of RF Cavities in High Magnetic Fields}} 
Accelerating gradients in excess of $50~\MV/\m$ in a $3$~T magnetic field have been demonstrated in the FNAL MuCool Test Area (MTA).

\noindent\underline{\it{Initial and 6D Ionization Cooling Designs and pioneering demonstration}}
\noindent Concepts were developed for Initial Cooling, and 6D Cooling with RF cavities operating in vacuum (VCC), including a variant on this design where the cavities were filled with gas used as discrete absorber (hybrid scheme), and a Helical Cooling Channel (HCC) design operating as a gas-filled channel. The recent results concerning the operation of RF cavities in high magnetic fields exceed the requirements for these lattices, thus opening up the possibility of further improvements.  Additionally, lower emittances can potentially be achieved by adding a final section utilizing High Temperature Superconductor (HTS) technology. The International Muon Ionization Cooling Experiment (MICE) \cite{MICE} collaboration, hosted at STFC Rutherford-Appleton Laboratory (RAL), undertook the task of characterizing the energy-loss and multiple scattering characteristics of muons in the momentum regime relevant for the construction of an ionization cooling channel ($\sim200$~MeV$/c$). It also demonstrated the technical feasibility of the key required components (magnets, RF, aborbers, etc). The MICE pioneering observation of the ionization-cooling of muon beams -- an emittance reduction of 6\% -- is well described by theory and simulations   \cite{Mohayai:2018rxn,Blackmore:2018mfr}. However, the experiment has been carried out with a transverse emittance of two orders of magnitude above the one required by a collider.\\
A test facility is required to demonstrate significant cooling in the transverse and longitudinal planes and to levels relevant for a collider scenario. Using a significant bunch charge will be essential to perform more complete measurements. This effort will help to select a cooling scheme among the various possible options and to gain confidence that it will achieve the targeted performance. In parallel, a feasibility study of the promising Parametric-resonance Ionization Cooling (PIC) \cite{Derbenev:2012dh} should be launched for its potential of improved cooling performance.

\noindent\underline{\it{Very High Field Solenoid Magnets}} are required for the final cooling to achieve low transverse beam emittances. A recent demonstration of a $32$~T, superconducting solenoid with $34~\mm$ cold bore has been carried out. This is in excess of the previously assumed magnetic field of $30$~T with a $25~\mm$ aperture used in final cooling scenario. This result together with the rapid improvements in HTS-based solenoids opens up the possibility of further optimization of the Final Cooling system design.

\noindent\underline{\it{Fast Acceleration to collider energies}} 
\noindent Since synchrotron radiation is not a limiting factor in accelerating muons to the TeV-scale, efficient multi-pass acceleration can be used for cost-effective collider facilities. At low energy, acceleration in Recirculating Linacs (RLA) and, from the $\sim100$~GeV-scale, a hybrid Rapid Cycling Synchrotron (RCS) concept where fixed field superconducting dipoles are interleaved with fast normal-conducting ramping magnets, have been identified as the most cost-effective solutions with efficient power consumption. Such a ring can be used to accelerate the muons to the final energy before they are injected into the high-field collider ring; hence its circumference can be much larger than the collider ring - in inverse proportion to the average field level achieved. Initial studies indicated that magnets meeting the requirements of $2$~T peak-to-peak operation at a minimum of $400~\Hz$ frequency are achievable, although further work is required to prepare a full-scale working prototype. A collider ring that also serves as the RCS has been suggested \cite{Neuffer:2018yof} and it is an option that requires further exploration.

\noindent\underline{\it{High-field and fast-ramping dipoles}} 
\noindent  High-field dipoles are essential in the collider ring to reach high energy and luminosity. The magnets have to be robust against the decay products of the muon beams that could quench them. High-field dipoles and fast-ramping dipoles with a large field variation are essential to limit the size and cost of the RCS. 

\subsection{Positron driver scheme}\label{sec:pods}

The proton driver muon source generates beams with relatively large emittance. This is due to the limited capacity to cool the initial large emittance muon beams by ionization in matter, where multiple scattering also occurs. Therefore relatively larger muon beam currents in the collider are mandatory to reach a high luminosity. When the muons decay while circulating in the collider ring, they generate a large background in the detector and also lead to significant radiation at the surface. This ultimately limits the energy and luminosity that can be reached in a muon collider.

The LEMMA scheme aims at overcoming this limitation by generating very small emittance muon beams, without any need for muon cooling.
This would allow for high luminosities with much smaller beam currents and consequently reduced detector background and surface radiation. This scheme strongly relaxes the limitations in energy and luminosity arising from the muon decay.

\subsubsection{Design status}

The Low Emittance Muon Accelerator (LEMMA) concept \cite{Antonelli:2015nla} is based on muon production from a $45$~GeV positron beam annihilating with the electrons of a target close to threshold for $\mu^{+}\mu^{-}$ pair creation, thus generating, without any cooling, muon beams with low enough transverse emittance for a high luminosity collider. The low muon conversion efficiency ($\sim 9\cdot10^{-8}$ muons per positron using a $3~\mm$ Beryllium target), motivated an extremely ambitious production scheme allowing to reach a high collider luminosity, similar to that of the MAP proposal. 
The initial design foresees a positron storage ring with an internal target, in order to allow multiple interactions of the positrons with the electrons at rest in the target.  
However this first layout has encountered several limiting difficulties.
One of them is the required intensity of the positron beam source. With a mean positron lifetime of 100 turns, a positron source providing $10^{16}~e^+/$sec allows for a flux of $10^{18}~e^+/$sec on target. This source is already two orders of magnitude more intense than the International Linear Collider one. 
Other issues are related to the instantaneous and the accumulated Peak Energy Deposition Density (PEDD) in the target for muon production; the impact of multiple passages through the target on the muon and positron beams; the recombination of multiple muon bunches before and after acceleration.
Presently an alternative accelerator complex design is under study, identifying the challenges within reach of the existing technology, and those requiring further innovation.  The final goal of this process is to come up with a revised start-to-end machine design and a well-defined R\&D roadmap.

\subsubsection{R\&D}

\noindent\underline{\it{Overall facility concept}} 
\noindent A redesign of this concept is needed to address the issues found in the initial scheme and to evaluate the achievable collider instantaneous luminosity, as limited by the muon production efficiency. The new scheme under study is still based on a $45$~GeV positron beam, but the positron bunches will be extracted to impinge on multiple targets in long straight sections with multiple IPs. This scheme could release the impact of the average power on the targets and also reduce the number of positrons needed from the source. Also, it would improve the luminosity by increasing the production yield with respect to the muon beam emittance. 

\noindent\underline{\it{Bunch stacking in the collider ring}} 
\noindent A major shortcoming in the initial scheme is that no realistic concept is presented for the combination of the bunches that are injected into the collider ring into a single bunch. A proposed scheme to overcome this in the longitudinal plane requires important R\&D for the related hardware in order to establish feasibility. Alternatively, one should explore the option to produce higher muon beam intensities already at the level of the source.

\noindent\underline{\it{Positron source and muon production target}} 
\noindent A novel design of the positron source could take advantage of synergies with other future collider studies. A more elaborate machine complex, exploiting both ring and linac assets should be envisaged, to produce and extract a high flux and low emittance positron beam, accelerated to 45 GeV, to impinge onto a system of multiple targets. The positron storage ring should be reoptimized toward longer positron lifetime with stronger focusing on the production target(s).
Systematic research on positron and muon production targets is required. For muon production an optimized choice of the target material is needed: beryllium, carbon composites or other solutions like liquid Lithium and liquid Hydrogen are under consideration. R\&D on specific target shapes and configurations to reduce the emittance of the produced muon beam are highly desirable to avoid the luminosity reduction. R\&D on the positron source could explore the use of Tungsten crystals, relying on the intense channeling radiation in axially oriented crystalline. Moreover R\&D on positron capture efficiency and the possibility of a multi-target scheme also for increasing the positron production are envisageable. The muon accumulation rings need to be designed with optimized matching onto the production target.

\noindent\underline{\it{Acceleration}} 
\noindent Depending on the source and collider ring design, two options can be explored.\\
 If a novel scheme can produce intense muon bunches at $15~\Hz$, the acceleration complex can be similar to the RCS-based proton driver scheme; the main difference being the lower bunch charge with smaller emittances. If the muon bunches are combined in the collider ring, higher repetition rate is required. In this case, the acceleration scheme will likely differ from the proton-driver case, relying on FFAG-RLA scenarios rather than a RCS.

\section{Backgrounds and Detector Studies}\label{sec:background}

The primary source of backgrounds in a muon collider detector is the $\mu\rightarrow e\nu\overline\nu$ decay of the muon beam. Detailed studies have been performed by the MAP group \cite{JINST} for a muon collider with center of mass energy of $1.5$~TeV. The electrons coming from the muon decays interact with the beamline components and generate electromagnetic showers. As a result, a large number of low-energy photons and soft neutrons may reach the accelerator elements and the detectors. Large heat deposition from the showers induced by decay electrons requires using large-aperture magnets to accommodate the thick high-Z absorbers needed to protect the superconducting coils. In particular intensive simulations are needed to design sophisticated radiation protection systems to shield the magnets apertures and interconnections. The lower intensity LEMMA source would enable thinner absorbers and smaller apertures.

Part of the background is generated by beam interactions many meters upstream of the interaction point. There is also a halo of decay electrons accompanying the beam to the interaction point. If there were no specifically designed shielding, this halo would produce an unacceptable level of electromagnetic background in the detectors. Any detector at a muon collider will then have to be designed with the backgrounds from muon decays taken into account. Detector studies performed in the context of the MAP effort \cite{JINST} show that the current approach to handle high detector backgrounds appear adequate to preserve the required physics capabilities.  However, these studies need to be extended, and to higher c.o.m. energies, to explore the full range of physics processes in detail and characterize the overall physics reach of any muon collider design.

Another important aspect should be taken into account: the decay neutrinos will produce a secondary radiation, with hadrons, muons and electrons traversing the earth that may constitute a radiological hazard. Even without detailed studies, \cite{JINST} suggests that the ultimate energy of a muon collider might be limited by neutrino radiation at ground level. It was evaluated that, following US federal limits, the ultimate center-of-mass colliding-beam energy could be limited to about $10$~TeV with a depth of $500~\m$. Some ideas (such as conveying neutrinos generated in the straight sections of the accelerator through a modest vacuum beam pipe) to mitigate the problem were proposed in \cite{JRS} and these require further development. In the case of the LEMMA scheme, the fact that muon beams from $e^+e^-$ collisions have a low emittance may allow high luminosity with a smaller number of muons per bunch. Therefore, a lower level of background is expected, enabling a higher center of mass energy before producing a radiological hazard. The radiation hazard is also an important concern in the use of existing tunnels that have straight insertions. They can increase the radiation at the locations where the straight extension of the tunnel would intersect the surface, if no countermeasures are used. 

\section{Implementations}

Muon-based facilities offer unique potential to provide a next generation of capabilities and world-leading experimental support, spanning physics at both the Intensity and Energy Frontiers. A complete picture has been identified within the framework of the Muon Accelerator Staging Study (MASS)  \cite{Delahaye:2015yxa}. 
A preliminary study of a center of mass $14$~TeV muon collider in the CERN LHC tunnel has recently been considered \cite{Neuffer:2018yof}. It leverages the existing CERN facilities, including the $26.7~\km$ circumference LHC tunnel and its injectors. Collisions at $14$~TeV center of mass energy of fundamental leptons could provide physics reach comparable to the interaction of the proton constituents in a $100$~TeV FCC-hh, but within the present CERN footprint.

The performance based on a proton driver scheme in a collider equipped with $16$~T magnets as developed for the FCC project and adapted for muon decay shielding would be especially attractive with a luminosity up to $10^{35}\cm^{-2}\sec^{-1}$ at the limit of on-site neutrino radiation.  A positron driver scheme with a potentially lower neutrino radiation could possibly extend further this energy range. The realistic performance and feasibility of such a scheme would have to be confirmed by a detailed feasibility study identifying the required R\&D to address its specific issues, especially the limitations from neutrino radiation on site and the compatibility of the existing CERN facilities with the muon decays. Synergies with the FCC developments in high magnetic field magnets and/or with the CLIC development of high accelerating gradients should be further explored.  

\section{Conclusions and recommendations }

Muon-based technology represents a unique opportunity for the future of high energy physics research: the multi-TeV energy domain exploration. 
The development of the challenging technologies for the frontier muon accelerators has shown enormous progress in addressing the feasibility of major technical issues with R\&D performed by international collaborations. In Europe, the reuse of existing facilities and infrastructure for a muon collider is of interest. In particular the implementation of a muon collider in the LHC tunnel appears promising, but detailed studies are required to establish feasibility, performance and cost of such a project. A set of recommendations listed below will allow to make the muon technology mature enough to be favorably considered as a candidate for high-energy facilities in the future.\\[4pt]
\noindent{\bf{Set-up an international collaboration}} to promote muon colliders and organize the effort on the development of both accelerators and detectors and to define the road-map towards a CDR by the next Strategy update. As demonstrated in past experiences, the resources needed are not negligible in terms of cost and manpower and this calls for a well-organized international effort. \\
For example, the MAP program required an yearly average of about \mbox{$10$M\$} and \mbox{$20$ FTE} staff/faculty in the $3$-year period 2012-2014.\\[3pt]
\noindent{\bf{Develop a muon collider concept based on the proton driver and considering the existing infrastructure.}} This includes the definition of the required R\&D program, based on previously achieved results, and covering the major issues such as cooling, acceleration, fast ramping magnets, detectors, \ldots.\\[3pt]
\noindent{\bf{Consolidate the positron driver scheme}} addressing specifically the target system, bunch combination scheme, beam emittance preservation, acceleration and collider ring issues.\\[3pt]
\noindent{\bf{Carry out the R\&D program toward the muon collider.}} Based on the progress of the proton-driver and positron-based approaches, develop hardware and research facilities as well as perform beam tests.\\[4pt]
Preparing and launching a conclusive R\&D program towards a multi-TeV muon collider is mandatory to explore this unique opportunity for high energy physics.
A well focused international effort is required in order to exploit existing key competences and to draw the roadmap of this challenging project.\\
The development of new technologies should happen in synergy with other accelerator projects. \\ 
Moreover, it could also enable novel mid-term experiments. 

\section*{Acknowledgements}
We are pleased to acknowledge the MAP, LEMMA and MICE collaborations for all the excellent developments already achieved on the promising but challenging muon technology, and for all the discussions and useful comments while preparing this document. We particularly thank M.~A.~Palmer.

\newpage

{\small{

}}
\end{document}